# Quantum Physics From Abstract to Laboratory Space I.
## Q-States Sustained by Partite Material Systems:
## Linking A⊕B and A⊗B domains via Entanglement

O. Tapia
Chemistry Ångström, Uppsala University, Sweden

The paper focuses on aspects of the measurement problem introducing quantum states (q-states) for measured and measuring systems. The link between non-interacting and interacting quantum systems is first look at. For two independent partite systems logical sums A⊕B stand for non-interacting q-systems; while a direct product space A⊗B gathers interacting states. However this latter should support physical q-states with base states that do not separately belong to either A nor B; the latter correspond to bridge states, namely entangled states that can perform as links (bridges) between A⊕B and A⊗B domains. Bridge states at laboratory space open possibilities to describe transport in quantized amounts of energy and angular momentum. These link bases sustain entanglements of different kinds. Interactions bring in quantized electromagnetic (em) fields. Matter sustained q-states entangled to em-sustained q-states open bridges to transport information between matter and radiation.

Immersed in a dominating classical worldview quantum physics initially imposed on quantum measurement outcomes an obvious *macroscopic* (classical physics) interpretive character. Yet at millennium turn the state of affairs including perceptions of both q-states and measurement concepts were gradually changing whence the distinction between abstract and laboratory domain states permits developing measurement frameworks that would actually sustain q-physical processes. Understanding this opening is an aim for this paper.

Abstract q-states are elements of Hilbert space. Physical q-states or laboratory space q-states are supported by given elementary materiality characteristic for the system. Measurement processes link physical states to particular partite q-states in measuring devices presented as semi-classic events. The classical occupation motif, particles occupying a q-state, is set aside now as a non-relevant feature concerning abstract space. Materiality is first left somehow aloof; once physical q-states are introduced, the only requirement is expressed by the idea of *presence* in the experimental (physical) domain; a sort of executive presence though not as entities (objects). Occupancy is no longer a "good quantum number" and at best it may be a useful *ansatz* in some other related semi-classic contexts. The latter stance displaces the representation viewpoint, viz. quantum states "wrapping" such materiality. Consequently elementary material substrate may sustain any number of q-states and quantum coherence would appear in a natural manner in laboratory setups as well.

As soon as one gets confortable with quantum concepts, e.g. coherent states, entanglement, spinor states, quantum operators (self-adjoint) in Hilbert space, the language would mold them to better reflect *quantumness* in uncovered results (e.g. spectra). The allegory used by Schrödinger to make clear the unsoundness of the standard quantum measurement theory was turned in derision. Yet, the idea of q-state as if it were representing an object misses the point; a q-state is not an object, it is an element of a mathematical abstract space always. The research task is to find out links between abstract and laboratory domains. And in particular this paper includes the chemical structure concept that in *abstract* quantum mechanics does not make sense while in the semi-classic approach it finds a central mechanical place: from the viewpoint adopted here it plays the role of bridge concept.

What usefulness such view might have? For one, quantum measurement becomes a physical process with no intervening observers. Registering such events also display the character of a physical process.

But first, there is need for a quantized view rather than a classical one. Especially when spectroscopic factors, ranging from low frequency radiation through UV and up to high-energy sources e.g. X-ray radiation, enter the theoretic descriptive field.

A photon-system, portrayed as an elementary energy and angular momentum carrier (executive presence), is non-separable. Thus, envisaging a one photon-state transport at the laboratory level will appear as q-interaction relating both elements. Quantum physics finds its birthplace at a fence between laboratory and abstract space as epitomized by Plank's discovery: energy *exchange* there takes place in finite amounts, namely, *quanta*. The exchange relates two q-elements.

In q-physics two sorts of interactions are discernable: Abstract q-operators coupling eigen-states in Hilbert space; and scattering operators that can also be presented by semi-classic operators parametrically dependent on laboratory space coordinates; for example a double slit device. A self-adjoint operator including configuration coordinates of the q-system and real space coordinates localizing the setup in real space is a typical semi-classic element. Besides, measurable energy/momentum exchange may in principle be symbolized by a q-event. In abstract space there is a change in the entangled states patterns constructed to render such processes. Photon states mediate energy/spin transfers with matter-sustained q-states; the latter would sustain implied information transfers.

In the following, an infinite dimensional vector space gathers *base states* and abstract q-states arranging sets of complex numbers (amplitudes) labeled by base states disposed as column vectors; constructing an application as a scalar product (inner product) of these two vectors one gets a function of the amplitudes that would define coherent q-states. Laboratory base states are thence sustained by the particular elementary materiality. They keep the mathematical form (labels) given to abstract space elements.

Each partite system presents its own base vectors recognized with a global label. The simple direct product of them permits recognizing possible asymptotic states after interaction. The q-interaction at laboratory level introduces concept of an after and a before, i.e. time as a material trace can be identified. Time runs between q-events. The abstract time parameter appears via unitary evolution operators with no operational relationship to the time managed with the help of q-events that may discretionally be used at laboratory level.

Entangled base states permits opening partite states to interaction answering back for amplitude shifts leading for example to spin triplet-triplet states transfer as shown below. At the laboratory level entering and/or exiting from entanglement requires spending energy quanta and these latter may show different frequencies as the case may be. In interaction radiation and matter sustained base states are non separable.

**Photonics**

Photonic frameworks [2,3] require electronic quantum states independent from classic degrees of freedom, e.g. instantaneous nuclear positions found in the Born-Oppenheimer model. [4] Elementary constituents *sustain* quantum states; electrons and nuclei support electronuclear (EN) q-states. The photonic mode of relating these levels takes away the representational role that early theory assigned to q-states.

On the other hand, difficulties encountered in developing quantum electrodynamics (QED) [5,6] are bypassed with the non-representational approach. These ones had prevented earlier construction of useful schemes wedding matter q-states and quantum electromagnetic radiation, e.g. Fock *photon-number* base states. Photonic quantum chemical scheme [2,3] makes it possible to give presence to Planck's fundamental quantum event derivable from his 1900 seminal work. Q-interaction and q-entanglement are thence implemented via *non-separable* base elements complementing the direct product bases associated to semi-classic models.

Semi-classic models in quantum physics and chemistry are certainly most useful; they help develop computing replicas in QED, molecular Quantum Mechanics that is naturally applicable to Quantum Technology designs. [6] However this class of models (without entanglement) sidesteps q-events that find expression in q-interactions, in particular, between measuring (probing) laboratory devices and the q-system under probe (measurement). Interaction between q-systems and measuring (probing) ones is embodied by q-events opening, on the one hand, one space to the other as it were. And on the other hand they prompt for descriptions of quantum scattering (q-scat).

Likewise, the concept of quantum state has been changing under impulses originated from developments in quantum technology; see papers in refs.[7-9] that address the theme from initially opposite directions; within the present approach these directions tend to converge; this result obtains once a physical framework let relationally *bridge* abstract and laboratory fields.

To achieve a particular model, electromagnetic (EM) and material purviews must both be quantized (q-fields); one gets Fock basis elements $|n_{\omega j}>$ for EM q-states and on the other hand $\{|ik(i)>\}_{i,k}$ for matter sustained basis.[3,4] These sets open a way to handle q-events thus allowing targeting of energy/angular momentum exchange epitomized by Planck's discovery; this quantum possibility was not available to earlier approaches based on molecular wavefunctions only. [10,11]

Measurement calls for entanglement /disentanglement as a resource to bridge probe/ probing ends thereby forcing introduction of entanglement events. Importantly, simultaneous variation in both q-fields implies non-separability at the level of base states prompting for q-interactions.

Considering this new situation, it is necessary to first move theory's grounds beyond semi-classic levels.[9,12-14] Moreover, as R.G. Newton clearly asserts:[13] a state vector does not describe the system itself; the idea of representation falls off as also remarked by us in refs.[9,14] Amplitudes not only serve to calculate probabilities but then also intervene in q-interaction presentations between matter sustained q-state and those associated to probing space.[14] Consequently, classical physics is not central to the measurement event to the extent that local realism can safely be set aside.[15,16] As illustrated below, a graded scheme permits linking abstract to laboratory spaces.

The paper takes ground from Quantum Electrodynamics ideas advanced by Dirac, Fock and Podolsky;[17] although used here to rather illustrate connections and subtleties of the relativistic limit and help to relate with non-relativistic models.

**Photon q-states: Fock space**

In Fock space photon number base states: $\{|n_\omega\rangle\}$ prompts for coherent states that come out naturally by using the basis set $\langle F\text{-}\omega \text{ basis}|$:

$$(|0_\omega\rangle \; |1_\omega\rangle \; |2_\omega\rangle \ldots |n_\omega\rangle \ldots) \rightarrow \langle F\text{-}\omega \text{ basis}| \qquad (1)$$

Connections are set up by creation ($\hat{a}\dagger |n\rangle = \sqrt{n+1} \; |n+1\rangle$) and annihilation ($\hat{a} |n\rangle = \sqrt{n} \; |n-1\rangle$) *operators* relating Fock base states; no particles are "created" or annihilated in a physical sense; in fact it is a *reckoning device* for abstract spaces. Note the "colored" vacuum base in (1) indicates it as a label (sub index) this is just information.

A particular q-state comes out as an infinite dimensional row vector with complex numbers (amplitudes) ordered according to the sequence found in the (1)-|basis⟩ and using basis labels as e.g.:

$$(C_0 \; C_1 \; C_2 \ldots C_n \ldots)^T \rightarrow |\text{q-state name}\rangle \qquad (2)$$

Families of photon q-states can be obtained as follows. Select a complex number $\alpha$ as eigenvalue of eq.: $\hat{a}|\alpha\rangle = \alpha|\alpha\rangle$, where $\langle\alpha|\hat{a}\dagger = \alpha^* \langle\alpha|$, so that one obtains amplitudes $C_0 = \alpha^0/\sqrt{0!}$, $C_1 = \alpha^1/\sqrt{1!}$, $C_2 = \alpha^2/\sqrt{2!}, \ldots, C_n = \alpha^n/\sqrt{n!}$, with a normalization factor, $A_0 = \exp(-|\alpha|^2/2)$; inserting these values in (2) a specific coherent q-state comes out as the special scalar product:

$$\langle F\text{-}\omega \text{ basis} | \text{q-state}(\alpha) \rangle := (|0_\omega\rangle \; |1_\omega\rangle \; |2_\omega\rangle \ldots |n_\omega\rangle \ldots)(C_0 \; C_1 \; C_2 \ldots C_n \ldots)^T$$
$$\rightarrow \Sigma_{k=0,1\ldots} C_k(\alpha) |n_{\omega,k}\rangle \qquad (3)$$

The term in a curly brackets $\{\ldots |i_\omega\rangle\langle j_\omega| \ldots\}$ relates to a density matrix operator. Given a quantum state, the expression $(C_0 \; C_1 \; C_2 \ldots C_n \ldots)\{\ldots |i_\omega\rangle\langle j_\omega| \ldots\}$ signals a transition operator vector associated to a particular quantum state. Shinning coherent light say at frequency $\omega$ only base states with non-zero amplitude value can originate (support) a transition response.

For, the commutator $[H, |i\rangle\otimes\langle j|]$ with eigen-states of Hamiltonian H leads to matrix elements $(\varepsilon_i - \varepsilon_j) |i\rangle\otimes\langle j|$. Applying this to a q-state one gets $(\varepsilon_i - \varepsilon_j) |i\rangle\otimes \langle j|\text{q-state}\rangle$ and therefore it picks up amplitude at the j-th base state so that it can modulate a response.

An energy quantum connects (so to speak) two energy level terms and Bohr's map: $(\varepsilon_i - \varepsilon_j) \rightarrow$ "quantity of EM-energy measured by the value of radiation frequency". Note that absolute value for energy levels is not meaningful; Bohr's map relates to energy level differences and in this sense become measurable as quantity of energy, namely, an energy quantum. *Eigen-values are not "observables" in themselves.*

This resource is offered by the quantum formalism. For a given q-state at probing it is apparent that exciting from a given base state (called a root state) requires the particular amplitude to be different from zero. This wedding of self-adjoint operators and the corresponding q-state opens a possibility to re-count abstract to laboratory quantities.

**Basic Photonic Scheme**
Direct products of EN-bases {|ik(i)>} and photon-base states enter as elements of an *extended* basis set, including now entangled terms that act as linking elements between separated matter and photon systems. A section that can be found in an infinite dimensional basis column vector of interest for a one-photon case reads: [2,3]

$$(\ldots |ik(i)\rangle \otimes |1_\omega\rangle \ldots |ik(i);1_\omega\rangle \otimes |0_\omega\rangle \ldots |i'k'(i');0_\omega\rangle \otimes |0_{\omega'}\rangle \ldots |i'k'(i')\rangle \otimes |0_{\omega'}\rangle \ldots)^T \rightarrow \langle\text{basis}| \qquad (4)$$

A particular q-state comes out as row vectors assembling complex numbers (amplitudes) ordered according to the sequence found in (4)-|basis> and using basis labels for the amplitudes one outlines particular abstract q-states:

$$(\ldots C_{ik(i)\otimes 1_\omega} \ldots C_{ik(i);1_\omega \otimes 0_\omega} \ldots C_{i'k'(i');0_\omega \otimes 0_{\omega'}} \ldots C_{i'k'(i')\otimes 0_{\omega'}} \ldots) \rightarrow |\text{q-state}\rangle \qquad (5)$$

Again, non-zero amplitudes control interactions with external probing devices while zero valued ones keep the set ordered. [3,4]

A particular q-state corresponds to an implicit scalar product: <basis|א|q-state>, where א-sign is there to prevent taking a simple (finite) sum; thus <basis |q-state> stands for the scalar product, <(4)|(5)>. Besides, *basis vectors* being an information resource they remain *fix once a particular model is chosen*. They are organized to outline possibilities associated to a particular materiality yet conserving their abstract character.

All terms in (4) form non-separable basis as they stand as possibilities associated to the quantum system and cannot be handled separately (independently); changes in q-states can happen via amplitude transformations engaging states like (5). No partitioning is yet included in the matter scheme that would allow for constituting subsystems spectral responses; inclusion will be assured if necessary later on. [2,3,14]

The element $|ik(i)\rangle \otimes |1_\omega\rangle$ from (4) shares a common origin (I-frame) with remaining basis elements otherwise it would stand for independent photon source and matter location. Imposing this caveat in (4) implies a constraint eliciting a sort of q-interaction. On the other hand the term $|ik(i);1_\omega\rangle \otimes |0_\omega\rangle$ indicates photon number depletion and would enter in presenting *entanglement of photon and matter field states;* note that there is no free photon state (quantum of EM field energy) available to be given up back if one were to startup probing at this level. The term stands as a photon-dressed base state (*photon-matter entangled state*) and provide executive presence to entanglement.

Next, take a base state e.g. $|i'k'(i');0_\omega\rangle \otimes |0_{\omega'}\rangle$ to imply an excited base state entangled to the photon field vacuum; while the simple direct product form $|i'k'(i')\rangle \otimes |0_{\omega'}\rangle$ opens matter sustained excited state interacting with a photon field vacuum. The actual amplitude value controls interaction (entanglement) process. [2,3]

If all indicated four amplitudes in (2) were different from zero, a coherent photon-matter q-state would obtain.

**Reading from q-states**

Consider a q-state amplitude changes say from $C_{ik(i) \otimes 1_\omega}=1$, $C_{i'k'(i');0_\omega \otimes 0_\omega}=0$ to $C_{ik(i) \otimes 1_\omega}=0$, $C_{i'k'(i');0_\omega}=1$ all other amplitudes being unchanged, one gets a state of photon/matter entanglement relating not to objects but q-state mutation possibly prompted by a q-process (not explicitly given) leading to *amplitudes relocations*. Consider a change:

From: $(\ldots 1_{ik(i) \otimes 1_\omega} \ldots 0_{ik(i);1_\omega \otimes 0_\omega} \ldots 0_{i'k'(i');0_\omega} \ldots 0_{i'k'(i') \otimes 0_\omega} \ldots )$ (6a)

To: $(\ldots 0_{ik(i) \otimes 1_\omega} \ldots 1_{ik(i);1_\omega \otimes 0_\omega} \ldots 0_{i'k'(i');0_\omega} \ldots 0_{i'k'(i') \otimes 0_\omega} \ldots )$ (6b)

Taken together (6a) and (6b) define a one-photon entrance channel seen from the matter elements viewpoint, from laboratory viewpoint: 1-photon state entanglement to sustaining matter. Materiality sustaining these q-states is necessarily *conserved* and never engaged in "filling" any energy eigenvalue.

Finite model linear superpositions such as: $(C_{6a} |(6a)\rangle + C_{6b} |(6b)\rangle)$ that normalized reads: $|C_{6a}|^2 + |C_{6b}|^2 = 1$ correspond to projected coherent states. These when apprehended globally, may possibly show finite lifetime; this latter engage the complete vector. This situation may result in links to q-events for displacements of energy and/or angular momentum. In fact, shifting amplitudes from-(6a)-to-(6b) the reading *inform* us a displacement of a unit angular momentum from a photon field to a matter sustained field; and consequently, selection rules would apply to couplings base states sustained by materiality.

Taken in the opposite direction, namely, from (6b) to (6a) one gets information that an emission mode opens as a possibility, the emitted state belongs to a limit that does not have its proper place in the space used to introduce $\langle(1)|(2)\rangle$ above. This simply means that abstract theory has to be supplemented to describe events, as one would expect. Old quantum mechanics completeness claims are not granted.

Once this caveat is understood one may go on. The amplitude at entangled base state in (6b') defeats propagation; at best this may act as initial state so that an external action that may be followed for instance by new steps such as amplitudes change from (6b') to (6c):

$(\ldots 0_{ik(i) \otimes 1_\omega} \ldots 1_{ik(i);1_\omega \otimes 0_\omega} \ldots 0_{i'k'(i');0_\omega \otimes 0_\omega} \ldots 0_{i'k'(i') \otimes 0_\omega} \ldots )$ (6b')

In eq.(6b') the entrance (emission) channel is closed; i.e., $0_{ik(i) \otimes 1_\omega}=0$. From here to state (6c) may result via coherent states, though for simplicity sake we show a pure state to introduce a simplifying language:

$(\ldots 0_{ik(i) \otimes 1_\omega} \ldots 0_{ik(i);1_\omega \otimes 0_\omega} \ldots 1_{i'k'(i');0_\omega \otimes 0_\omega} \ldots 0_{i'k'(i') \otimes 0_\omega} \ldots )$ (6c)

This latter stands as an excited state *entangled* to a *ω-vacuum*.

From a semi-classic stand the possibility of a photon state absorption (6c) starting from (6a) is apparent. Possibilities for having registered histories are brought in; the use of laser sources might help; physical examples are analyzed in Ref. [14].

Remind: elementary materiality does not occupy individual base states (energy levels) so that all basis positions remain open as possibilities modulated by amplitudes; these ones turn out operative whenever a time evolution process starts up or a laboratory event prepared by an experimenter is flashed. [14]

Why don't we use the propensity idea? Simply because the photonic quantum chemical scheme is not representational, nothing is said on the elementary material constituents except that they must always be present (sustaining as an abstract configuration space would do). This is a much weaker "metaphysical" assumption concerning material reality than the one found in standard von Neumann QM.

**Abstract and laboratory spaces: Linking systems**

In terms of information data the bridge is clearly identified (see above). Special relativity theory (SRT) framework (I-frame) opens a connection (sort of gangplank) between abstract and laboratory domain. It helps introduce *configuration spaces* with dimension defined by *the number* of classical degrees of freedom, i.e. a dimensionless number. Two I-frames may define relative origins and orientation required by SRT.

There is no absolute space or absolute time.

**Role of configuration space**

Abstract configuration space: $\mathbf{x} \rightarrow (\mathbf{x}_1,\ldots,\mathbf{x}_n)$ collects information on the number 3n of classic degrees of freedom; these n–tuples support a linear vector space over the field of real numbers. In abstract space these numbers do not refer to "particle" properties, they enter as labels in, e.g. rigged Hilbert spaces either in configuration $\{|\mathbf{x}\rangle\}$ or reciprocal $\{|\mathbf{k}\rangle\}$ spaces. These base states would bridge (link) abstract q-states $\{|\psi\rangle\}$ to laboratory sets via projected states $\langle\mathbf{x}|\psi\rangle$ or $\langle\mathbf{k}|\phi\rangle$; or simply *wavefunctions*: $\psi(\mathbf{x})$ or $\phi(\mathbf{k})$ so the link is via I-frames located in laboratory space (real space). $\psi(\mathbf{x})$ or $\phi(\mathbf{k})$ are complex functions over real numbers support. Bases functions such as $\langle\mathbf{x}|\mathbf{k}\rangle$ or $\langle\mathbf{k}|\mathbf{x}\rangle$ have the form $\exp(i\mathbf{k}\cdot\mathbf{x})$ or $\exp(-i\mathbf{k}\cdot\mathbf{x})$ and used with measures $\mathbf{dx}$ or $\mathbf{dk}$ that introduce geometry elements. In $\mathbf{x}$-space $\mathbf{k}$ plays the role of a quantum number albeit a continuous one; $A(\mathbf{x})$ wave packet state in $\mathbf{x}$-space: $\int A(\mathbf{x})\exp(i\mathbf{k}\cdot\mathbf{x})\mathbf{dx} \Rightarrow B(\mathbf{k})$.

In $\mathbf{k}$-space wave packets read: $\int B(\mathbf{k})\exp(-i\mathbf{k}\cdot\mathbf{x})\mathbf{dk} \Rightarrow A(\mathbf{x})$ maps to a wave packet state in $\mathbf{k}$-space. Configuration spaces are just collections of real numbers, there is no special assigned meaning, they correspond to a type of abstract space.

**Time-frequency regime**

Bases functions such as $\langle t|\omega\rangle$ or $\langle\omega|t\rangle$ have the form $\exp(i\omega t)$ or $\exp(-i\omega t)$ with measures $d\omega$ or $dt$, respectively. In t-space $\omega$ play the role of a quantum number albeit a continuous one; $A(t)$ gives the wave packet state in t-space: $\int A(t)\exp(i\omega t)dt \Rightarrow B(\omega)$. And, in $\omega$-space wave packets read: $\int B(\omega)\exp(-i\omega t)\,d\omega \Rightarrow A(t)$ maps to a wave packet state in $\omega$-space.

To open for energy-time regime Planck's constant h/2π ($\hbar$) is required: It is a dimension based conditions not a quantization constraint.

**Semi-classic models**
Here coordinates refer to particle positions, e.g. electrons and/or nuclei. The numbers {**x**} are thus loaded with a meaning (that is to be classical particle positions). It goes without saying: it is not required by abstract mathematical structures, because no classical representation is sought. Labels 1-to-n can only be traced to particular classical material elements; this choice allows use of invariant ordered configuration space; for identical elements a second sub-index facilitates handling of permutations symmetries, see POL.[1] Quantum degrees of freedom would enter as new quantum numbers (labels, e.g. 2- and 4-spinors). Note the possibility to label the wavefunction with a global spin quantum number. At any rate, components' spinors cannot be handled separately as if a "particle" for example had spin α or spin β case S=1/2. You always work with a full spinor so that any rotation can properly be presented.

I-frames can distinguish internal and external q-states: a collection of n-partites referred to independent systems can be related to a one I-frame used to defining the n-tuple. In what follows we focus on quantum numbers for the internal as well as for quantized electromagnetic (EM) systems where a source defines an origin or else signals a target for emission case. Including quantum numbers covering spectral products and intermediate excited (transition) states in eq. (1), schemes obtain where chemical processes become explicit via *amplitude changes* of states type-(2). [2,4-8]

In practice, and it is the stance of the paper, semi-classic and full quantum physical schemes complement each other.[2,5-8] The photonic approach projected with the help of configuration space leads to a different view of chemical processes. To help appraisal of q-state notion, consider an experimental double-slit experiment [16] as it can be viewed from the present framework.

**Quantum physics of Tonomura double-slit experiment**
Sensing the interplay between abstract and laboratory levels, this experiment, using electrons as carriers, touches key aspects of quantum physical foundations: its clarification provides a new awareness of quantum states.

Electron states are basic elements that supplemented by detection devices where q-events are registered provide uncovering of associated q-states. Entangled states mediate possible q-energy transfers to-or-from one subsystem to the other; location is not determined by the electron q-state, only changes of can be given some space time characters. Tonomura developed technology so that the equivalent of one-electron-at-a-time was present in the region supporting a device comparable to a double slit.[16] Actually, the statement "only *one electron* at a time came" must be attuned: *only one elementary materiality was present at a time to sustain the q-state*. This latter element is the relevant component for the quantum theoretical analysis; only its executive presence is required.

In other words, a q-state *interacts* with the double-slit that technically is given by an interaction operator conveying relevant geometric (laboratory) information. It is neither a particle nor a path that are the relevant features. It is the q-interaction that is to be explored: first, from an abstract standpoint thereby skimming all possibilities; and there after semi-classic elements enter to complete the description. Quantum states are first handed in Hilbert-Fock space, not in real (laboratory) space, then interaction with a double slit device is given a q-operator form; these interaction centers are associated to scattered q-states eliciting sets of possibilities. Once you understand that a full set of possibilities is involved, the interference pattern is a natural mathematic result. Questions such as: which way took the photon are meaningless.

Thus, in abstract space all possibilities would be included. In this case the result is a q-state signaling an interference pattern. This situation (in abstract space) can be set in correspondence (linked) to simultaneous interaction at the double slit device with *one and the same incident q-state*. The experimental setup grants this sameness. Thus, interposing a detecting surface beyond location of the double slit, and in agreement to (semi-classic) calculations, an *interference pattern* should emerge from actual q-events; and it does as predicted with q-physical tools. Experimentally the case presents as follows.

Detection of electron states one-by-one generates apparent random clicks during initial collection of spots (clicks). However, according to present perspective (sustainment), any two q-events will be independently correlated via the final quantum state containing full information on q-interactions with the double slit. Yet, any sustaining material would be a "carrier" for the *same* q-state though *prediction*, in a classical physics sense, of spot localization is not legitimate. No independent particles model can do it; we are thence in presence of q-interactions at the double-slit (or its equivalent) yielding q-states sensing a final interference scheme. The abstract interference pattern will be there, first as calculation possibility, and, until enough incoming q-events impinge with the recording device, an *image* of that interference would slowly emerge.

It is a q-event that can engender a "click"; this one would be a mimicker of a quantum energy/momentum transfer; a sensitive surface that records events, initially after collecting a few of them, would appear *as if they were random in location*. Actually, early events do *appear* as being quite randomly distributed (noticed by Tonomura [16]); this is the impression at least and suggests that a too positivistic *interpretation* misses the physics encapsulated in a q-event. Gathering q-events in a separate device once all of them are simultaneously exposed a clear (interference) image will emerge as a result of the underlying q-state. *The theory cannot predict separated events locations*. Such is the character of quantum physics.

**1-photon-state initiating quantum physical processes**
What one-photon scheme case does achieve? Some examples are given henceforward.

Consider:
i) Base state $|i=0> \otimes |1_\omega>$ features spin 1 from the photon base state and possible interaction with electronic ground state, i=0. The classic side of a q-interaction;
ii) Entanglement base element $|i=0;1_\omega> \otimes |0_\omega>$ displays spin 1 this time sustained by the entangled photon-matter term;
iii) First electronic excited state $|i=1, S=0, L=1; 0_\omega> \otimes |0_\omega>$ angular momentum (AM) lies at L=1;

iv) Lowest spin triplet $|i=2,T\rangle \otimes |0_\omega\rangle$; it shows angular momentum 2 namely L=1, S=1. The updated base vector looks as (6):

$$(|i=0\rangle \otimes |1_\omega\rangle \ldots |i=0;1_\omega\rangle \otimes |0_\omega\rangle \ldots |i=1,S=0;0_\omega\rangle \otimes |0_\omega\rangle \ldots |i=1,S=0\rangle \otimes |0_\omega\rangle \ldots |i=2,T\rangle \otimes |0_\omega\rangle \ldots)^T \quad (6)$$

These elements open to states sustained by the elementary materiality. Q-base state (6) is a unity, a one-partite state, where the first entry would mediate the connection with a possible bi-partite base associated to e.g. free photon and elementary materiality. It is beyond a q-theory to represent (as it were) the link, only a q-interaction would relate these two worlds.

The energy associated to the first four terms in (6) is the same; so, there is no jump when one sees (6) as a unity; only if one focus on one partner (materiality sustained states for instance) there is a change of both energy and AM. Observe: a photon-matter entangled basis cannot be separated; it is a "unit" though not an object. A way out to the left side (propagation) would change amplitudes so that the response from the entangled basis changes into a response from amplitudes at e.g. $|i=1,S=0\rangle \otimes |0_\omega\rangle$ that is an "isolated" excited state. If one eliminates all entanglements of this kind then a standard basis set for one EN Hilbert space obtains. But even in this case the system can only display coherent states. The inclusion of photon basis, in the way shown so far, corresponds to a physical Hilbert space prepared to bridge both systems. Let explore some possibilities the framework offers.

**Opening access to spin triplet states: Optically assisted pathway-model**

Amplitude at $S_1$ that can be directly open from a ground closed shell state ($S_0$) while spin triplet base ($T_1$) state features zero amplitude value because $S_0 \to T_1$ and $S_1 \to T_1$ are both forbidden transitions by AM conservation.

A possible activation takes on a "route" similar to the optically activated zero field magnetic resonance phenomena.[19,20]

Thus, $(C_{i=0 \otimes 1_\omega} \ldots C_{i=0;1_\omega \otimes 0_\omega} \ldots 0_{1;0_\omega} \ldots 0_{1 \otimes 0_\omega} \ldots 0_{2,\otimes 0_\omega} \ldots)$ links to a coherent state that might connect to an entrance channel corresponding to q-states covering other spectral sectors. This case heralds a photon state entangled with ground state basis. Both amplitudes being different from zero simultaneously; energy conserves if relation $|C_{i=0 \otimes 1_\omega}|^2 + |C_{i=0;1_\omega \otimes 0_\omega}|^2 = 1$ holds. The very activation process appears sustained by entanglement of one-photon and ground states.

The situation signaled above permits apprehending grounds for an elastic scattering description: targeting ingoing state $(1_{i=0 \otimes 1_\omega} \ldots 0_{i=0;1_\omega \otimes 0_\omega} \ldots)$, entanglement with q-state $(0_{i=0 \otimes 1_\omega} \ldots 1_{i=0;1_\omega \otimes 0_\omega} \ldots)$ and scattering situation via $(\exp(-i\mathbf{k'}\cdot\mathbf{r})1_{i=0 \otimes 1_\omega} \ldots 0_{i=0;1_\omega \otimes 0_\omega} \ldots)$ where $\mathbf{k'}$ signals any direction away the I-frame of targeted state. Note that amplitude term $\exp(-i\mathbf{k'}\cdot\mathbf{r})1_{i=0 \otimes 1_\omega}$ cannot be taken in isolation; otherwise, one re-introduces a particle concept. The coherent state speaks of possible cases, though not of representing (whatever the case might be). *Quantum formalism takes over all possibilities*. And this is what it makes so different from classical physics (mechanics) ones.

Now we can examine propagation from state (6') below over base states made accessible by a *second* one-photon interaction. Focus attention on fluorescent-like states:

$$(0_{i=0\otimes 1_\omega}\ldots 0_{i=0;1_\omega 0_\omega}\ldots 1_{1;0_\omega}\ldots 0_{1\otimes 0_\omega}\ldots 0_{2,\otimes 0_\omega}\ldots 0\ldots). \qquad (6')$$

From here (6'), triplet state activation may follow a course e.g.: 1) Information-injection of $S_1$-$T_1$ energy gap via a supplementary photon state $|1_{\omega S1\text{-}T1}>$; and 2) entanglement suggested by state (7) below identified by amplitude $C_{1*;0\omega\otimes 1\omega ST}$:

$$(0_{i=0\otimes 1_\omega}\ldots 0_{i=0;1_\omega\otimes 0_\omega\otimes 1_\omega ST}\ldots C_{1;0_\omega\otimes 0_\omega ST}\ldots C_{1*;0_\omega\otimes 1_\omega ST}\ldots 0_{2,\otimes 0_\omega\otimes 0_\omega ST}\ldots) \qquad (7)$$

Note the addition of the excited state base state $|1*;0\omega\otimes 1\omega_{ST}>\otimes|0_\omega>$; by construction it has two pieces of information i.e. energy level above the first excited state ($S_1$) and the energy level when measured from ground state ($T_0$) displays the equivalent of $2\hbar\omega_{ST}$ so that the triplet base state label now gains new information on the $S_1$-$T_1$ channel. Energy conservation demands inclusion of a second external photon state $|1\omega_{ST}>$; this one would "harvest" an assisted consecutive two photon emission; it would result in the triplet state opening with a non-zero value for the amplitude, i.e.:

$$(0_{i=0\otimes 1_\omega}\ldots 0_{i=0;1_\omega\otimes 0_\omega\otimes 1_\omega ST}\ldots 0_{1;0_\omega\otimes 0_\omega ST}\ldots 0_{1*;0_\omega\otimes 1_\omega ST}\ldots 1_{2,\otimes 0_\omega\otimes 0_\omega ST}\ldots 0\ldots) \qquad (7')$$

The triplet base state shows $L=1$, $M_L=0$ sustained by space anti-symmetric term and spin state with q-number equal $S=1$, $M_S=0$. The effective production would consume two units of AM measured here as 2 photons; the required AM quantity were taken from external photon fields. This is an equivalent of spin-orbit coupling in the semi-classic scheme.

Observe that a direct one-photon activation from singlet ground state $S_0$ to the triplet is not possible due to AM-conservation rules *unless the photon state also display space (L>1) AM*.

Once the activation channel displays non-zero amplitude at state (7) only the first excited state $S_1$ would act as root state for supplementary electronic excitation events. A *non-rotating* wave model supplies energy at $2\hbar\omega_{ST}$ that add label information at $0_{2,T\otimes 0_\omega}$ so that it can play the role of a dressed vacuum, i.e.: $0_{2,T\otimes 0_\omega\otimes 0_\omega ST}$. Standard rotating frame model is not useful in this context it is too classic.

Thus, shine a *second* ST-gap photon state $|1_{\omega ST}>$ at (7), to prompt for the cascade process noted above; this would end up with non-zero amplitude at $C_{|i=2,S=1>\otimes|0_\omega>\otimes|0_\omega ST>}$ these states remain implicit in the notation of (7) that now we change into the more explicit end result:

$$(\ldots 0_{i\otimes 1_\omega}\ldots 0_{i;1_\omega\otimes 0_\omega}\ldots 0_{i+1;0_\omega\otimes}\ldots 1_{i=2,T\otimes 0_\omega 0_\omega ST}\ldots) \qquad (8)$$

The information brought up by injection is registered by the second ST-photon gap state that would add amplitude shifting from $0_{i+2,T\otimes 0_\omega 0_\omega ST}$ into $1_{i+2,T\otimes 0_\omega 0_\omega ST}$; the operation shares a flavor of information supplement via the label. In other words, chemical and photo-physical processes starting at a triplet as root state can now begin. The ST-gap two photons pay for two units of angular momentum required by the "transfer" from the spin-singlet electronic excited state. The space part corresponds to changing from $L=0$ to $L=1$. The supporting elementary materiality remains unchanged in numbers.

Note that one ST-gap photon state could be radiated thereby possibly acting as a catalyst to activation of the triplet state reflected in (8).

Note, the elementary event that may lead to a transfer of one energy quantum could have taken (6') as portal state. *Mutatis mutandis* this type of state may also act as possible portal for photon emission. At any rate, *information circulates expressed via q-state amplitude changes*.

**Photon up-conversion: Triplet-triplet 2-photon interactions**

Up-conversion processes involving (spin) triplet states play key roles in organic light-emitting diodes (LED). Here, we present a photonic description in the style of optically assisted opening of spin triplet states.

Consider a bi-partite (dimer) with two separate sets of elementary materials sustaining equivalent spectra that, moreover, could independently be controlled. Each partite can be monitored following its particular I-frame.

Applying the procedure of the preceding section so that each dimer element (partite) is set at a triplet + triplet state Cf. Eq.(8)). The model assumes that each partite crystal site (or molecule) for instance sustains the respective I-frame thereby allowing for potential *control of relative real space distance between internal states*. To fuse both quantum internal spaces interaction would require entanglement. Thus the weakly interacting pair would appear as a linear combination of site states (LCSS):

$$1/\sqrt{2}\ [(\ldots 1_{2,\otimes 0_\omega \otimes 0_\omega ST}\ldots)_1 \pm (\ldots 1_{2,\otimes 0_\omega \otimes 0_\omega ST}\ldots)_2] \rightarrow 1/\sqrt{2}\ (|T_1>_1 \pm |T_1>_2) = |\pm> \qquad (9)$$

This equation looks like LCAO (linear combination of atomic orbitals) but here via entanglement mediation ("mechanism") the elementary material is expanded and the local q-states $(\ldots 1_{2,\otimes 0_\omega \otimes 0_\omega ST}\ldots)_{i=1,2}$ are disposed (included) in the extended Hilbert-Fock space.

The energy level for $|T_o>$ defines a zero energy level of its kind (triplet ground state) measured from $S_o$ ground state. Thus, under resonance condition elicited by (9) even a weak q-interaction would mix $|+>$ and $|->$ q-states thereby leading to $2E_{T1}$ at one site $|2T_1>$ and a triplet ground state $|T_o>$ at the other site. The interesting result: characteristic frequency $\omega_{ToT1}$ is then doubled at one site $2\omega_{ToT1}$ and formally zeroed (vacuum triplet state) at the other site. All these being *one and the same q-process*.

From the base state $|2T_1>$ several possibilities become open. Remember that two triplets combine to give: S=0,1 or 2 spin manifolds. Among the new possibilities there will be an excited singlet state with energy level about $2E_{T1} \rightarrow E_{S1^*}$. The energy gap with-respect-to the singlet ground state presents frequency $\omega'$ much larger that the one for the first singlet excited state. This is then one of the resources for the so-called photon up-conversion.

Note that the reading of q-states first, it does not produce what is being signaled, second it inform us only. There are no physical actions. Only a q-event would produce a physical mark possibly measurable, or at least opened to recording processes. See below for further cases.

**Probe thru X-ray and higher frequency photon states**

Recently the extension of laser techniques to coherent beams into the X-ray region of the EM spectrum to *tabletop equipment* opens opportunities for applications in exciting research fields. [21] The X-ray q-states generated represent a quantum coherent mode of the Roentgen X-ray tube in the soft X-ray region.

The energy quanta associated to this type of probe might be found well above first or second, ionization limit of materiality sustaining the processes.

Consider a generic circumstance: First step in the interaction with matter shares the same pattern, which corresponds always to entanglement thence a number of possibilities (non-radiative changes) are accessible until reaching at a possibly ionization channel such as:

$$(|i=0\ k(0)\rangle \otimes |1_\omega\rangle \ldots |i=0\ k(0);1_\omega\rangle \otimes |0_\omega\rangle \ldots |i=n^*\ k(n^*);0_\omega\rangle \otimes |0_\omega\rangle \ldots |n\ k(n)\rangle \otimes |0_\omega\rangle \ldots)^T \quad (10)$$

The energy labels n are always larger that n*; they may belong to a continuum. A generic quantum state takes on the form:

$$(\ C_{|i=0\ k(0)\rangle \otimes |1_\omega\rangle} \ldots C_{|i=0\ k(0);1_\omega\rangle \otimes |0_\omega\rangle} \ldots C_{|i=n^*\ k(n^*);0_\omega\rangle \otimes |0_\omega\rangle} \ldots C_{|n\ k(n)\rangle \otimes |0_\omega\rangle} \ldots)$$

The first interaction slot opens the material system via (12) namely photon –matter fields entanglement:

$$(\ C_{|i=0\ k(0)\rangle \otimes |1_\omega\rangle} \ldots C_{|i=0\ k(0);1_\omega\rangle \otimes |0_\omega\rangle} \ldots 0_{|i=n^*\ k(n^*);0_\omega\rangle \otimes |0_\omega\rangle} \ldots 0_{|n\ k(n)\rangle \otimes |0_\omega\rangle} \ldots) \quad (11)$$

These are the entangled states with energy equivalent to the incoming photon field. The initial slot being: $(1_{|i=0\ k(0)\rangle \otimes |1_\omega\rangle} \ldots 0_{|i=0\ k(0);1_\omega\rangle \otimes |0_\omega\rangle} \ldots 0_{|i=n^*\ k(n^*);0_\omega\rangle \otimes |0_\omega\rangle} \ldots 0_{|n\ k(n)\rangle \otimes |0_\omega\rangle} \ldots)$.

The form (11) of this q-state yet entangled via coherence with the photon vacuum requires "flowing in" the mater field space to become a special q-state e.g.:

$$(0_{|i=0\ k(0)\rangle \otimes |1_\omega\rangle} \ldots 1_{|i=0\ k(0);1_\omega\rangle \otimes |0_\omega\rangle} \ldots 0_{|i=n^*\ k(n^*);0_\omega\rangle \otimes |0_\omega\rangle} \ldots 0_{|n\ k(n)\rangle \otimes |0_\omega\rangle} \ldots). \quad (12)$$

Opening probing channel $0_{|i=0\ k(0)\rangle \otimes |1_\omega\rangle}$ now closed. Straight away, if a low frequency field (e.g. microwave or radio waves) acts on the system one can expect time dependence first at entrance channel amplitudes imposing time dependence to (11):

$$(\ C_{|i=0\ k(0)\rangle \otimes |1_\omega\rangle}(t) \ldots C_{|i=0\ k(0);1_\omega\rangle \otimes |0_\omega\rangle}(t) \ldots 0_{|i=n^*\ k(n^*);0_\omega\rangle \otimes |0_\omega\rangle} \ldots 0_{|n\ k(n)\rangle \otimes |0_\omega\rangle} \ldots).$$

Since the energy level $\varepsilon_{n\ |k(n)\rangle \otimes |0_\omega\rangle}$ features the same energy that $\varepsilon_{|i=0\ k(0);1_\omega\rangle \otimes |0_\omega\rangle}$ they can be coupled to low-frequency photon-fields.

There is a large energy gap between activated states and the rest. And observe that the lowest energy level would shift so that the coherent state forbids direct access to any external radiation. Thus, in the present model there are no free electron states in the continuum. The *ionization event (if any) would be mediated by an entangled state linking abstract to external space*. This latter one we have not yet demarcated.

Time dependence of (12')-type may be a source of electron states in the continuum:

$$(\ 0_{|i=0\ k(0)\rangle \otimes |1_\omega\rangle}(t) \ldots 0_{|i=0\ k(0);1_\omega\rangle \otimes |0_\omega\rangle}(t) \ldots 0_{|i=n^*\ k(n^*);0_\omega\rangle \otimes |0_\omega\rangle} \ldots 1_{|n\ k(n)\rangle \otimes |0_\omega\rangle}(t) \ldots) \quad (12')$$

State (12') could gate for a sufficiently large n a free-electron state sustained by a proper I-frame and positive-ion-state partite. This requires of a q-event sharing some similarity to "tunneling" via amplitudes.

And back again to (10) that has to be adjusted in order to include electron spin base states. The electron states correspond to 2-spinors. Each electronic quantum number includes spin S and a projection along an arbitrary direction $M_S$; the dimension $2S+1$ gives the range $-S \leq M_S \leq +S$. For $S=1/2$, two orthogonal base states are: $|\beta\rangle \to |S=1/2\ M_S=-1/2|\rangle$ and $|\alpha\rangle \to |S=1/2\ M_S=+1/2|\rangle$; these base states are sustained by two elementary materials (two electrons). For the atomic K-shell the electronic configuration reads: $1s^2$. For (12') there will be one electron at state $|1s^2\rangle$ and $|1s^1 1s^0\rangle$ where $1s^0$ indicates a hole at the K-shell. The state (12'') must be reshaped:

$$(1\alpha_{|i=0\ k(0)\rangle \otimes |1\omega\rangle}(t)\ldots 0_{|i=0\ k(0);1\omega\rangle \otimes |0\omega\rangle}\ldots 0_{|i=n^*\ k(n^*);0\omega\rangle \otimes |0\omega\rangle}\ldots 1\beta_{|n\ k(n)\rangle \otimes |0\omega\rangle}(t)\ldots)$$

$\alpha$ spinor (1 0) and $\beta$ spinor (0 1) so in this form they are correlated.

For an ionized state corresponding to release of one electron state measurable at lab-space the corresponding q-event is implicit (not computable). With this caveat the one-electron level is assigned an arbitrary spin ½ formally as:

$$((C_\alpha\ C_\beta)_{|i=0\ k(0)\rangle \otimes |1\omega\rangle}\ldots 0_{|i=0\ k(0);1\omega\rangle \otimes |0\omega\rangle}\ldots 0_{|i=n^*\ k(n^*);0\omega\rangle \otimes |0\omega\rangle}\ldots 0\ldots)$$

This form may signal an un-polarized spin state and an implicit hole at electronic level i=0. The precursor to the emitted electron state must be a p-state to compensate spin one taken away by the ionizing excitation. As low-frequency probes act on this state it may lead to coherent-like ones propagating in time:

$$((C_\alpha\ C_\beta)(t)_{|i=0\ k(0)\rangle \otimes |1\omega\rangle}\ldots C_{|i=0\ k(0);1\omega\rangle \otimes |0\omega\rangle}(t)\ldots C_{|i=n^*\ k(n^*);0\omega\rangle \otimes |0\omega\rangle}(t)\ldots 0\ldots) \qquad (13)$$

Once a laboratory event happens this class of path become a dead end and "traded" by a bi-partite situation (comparable to a laboratory electron release).

Intra q-state electronic transitions become possible now sustained by an ionic partite. For spherical symmetric hole-state any target state associated to angular momentum equals to 1 would present an allowed transition and consequently one envisages a corresponding photon emission. The source of this process (root state) displays symmetry l=1 and the corresponding target states requires an energy release with symmetry touching either a state l=0 (s-state) or l=2 (d-state); n≥2. If energy is sufficient to put the state in the continuum, a secondary electron might be detected. This is known as Auger electron and the full process as an *Auger process*.

Note that if the energy recovered by annihilating a hole-state does not suffice to ionize again the site, one would have a possibility to let propagate energy internally or to other sites if one handles solid-state cases.

In solid-state cases, the energy released in filling a hole-state may be worn out to set up an electron state in the conduction band. We may have thence a possibility for electric charge flux in the elementary material sustaining the q-states.

**Chemistry from a photonic quantum physical perspective**
The concept of molecular structure, chemical structure, underlies most of descriptive chemical phenomenology. In Quantum Chemistry the corresponding algorithms (e.g. based in Born-Oppenheimer model) lead to electronic wavefunctions depending parametrically on nuclear positions and typified by nodal planes patterns (NPP). The NPPs yield a partitioning of the nuclear space useful as communication tool. The point is that both, semi-classic and photonic frameworks address the same quantum system and, consequently, it would be possible to develop "ties" (bridge, gangplanks) in order to get a more comprehensive understanding of chemical processes, specially in this age of tabletop lasers.[21] The approaches have to be complementary, not exclusive of, which is the temper of our work.

The basis set for the photonic scheme has the form given in eq.(4). Thus, in principle, a huge number of energy states can come into a chemical horizon; many of them almost degenerate though with zero-value amplitudes at a given time.

Chemistry, among other things, is about forming/breaking bonds or in present language changing the number of partite states. From a one- to two- partite situation the abstract quantum description must relate domains accepting different numbers of I-frames. These latter are essential to describe laboratory events: e.g. in the scattering of two partite elements. Care must be exercised because domains might not be commensurate. The bi- and multi-partite system permits their laboratory space localization, which is a communication advantage. This is not a Hilbert-Fock space character in spite of the fact that semi-classic schemes mix them up.

*A q-event may thus tie (link) both spaces under specific circumstances*, e.g. laboratory probing. Consider a general piece-wise basis:

$$(|1g2g\rangle \otimes |1_\omega\rangle \ldots |1g2g;1_\omega\rangle \otimes |0_\omega\rangle \ldots |1g2e;0_\omega\rangle \otimes |0_\omega\rangle \ldots |1e2g;0_\omega\rangle \otimes |0_\omega\rangle \ldots)^T \rightarrow \langle\text{pw-basis}| \quad (14)$$

Both subsystems stand at the "door" of an entanglement event that quantum states like (15) below may fix in their traits. The second slot stands for an entangled base state case so that the photon field energy appears now "smeared" into the fix material constituent.

For quantum states such as: $(C_{|1g2g\rangle \otimes |1_\omega\rangle} \ldots C_{|1g2g;1_\omega\rangle \otimes |0_\omega\rangle} \ldots 0_{|1g2e;0_\omega\rangle \otimes |0_\omega\rangle} \ldots 0_{|1e2g;0_\omega\rangle \otimes |0_\omega\rangle} \ldots)$, the amplitudes cover radiation-matter interaction as if it were a portal gate that eventually could be used to open the system and trap energy by closing off with $C_{|1g2g\rangle \otimes |1_\omega\rangle}=0$. Though entanglement or taking a time-dependence driving in opposite direction one would get: $1_{|1g2g\rangle \otimes |1_\omega\rangle}$ and $0_{|1g2g;1_\omega\rangle \otimes |0_\omega\rangle}$ where amplitudes in one sense or the other refers to a one photon state that eventually can be emitted *if a proper q-event were to happen*. Classical predictability is thereby lost.

When amplitudes show a continuing propagation the state below signals families of excited states: $(0_{|1g2g\rangle \otimes |1_\omega\rangle} \ldots 0_{|1g2g;1_\omega\rangle \otimes |0_\omega\rangle} \ldots C_{|1g2e;0_\omega\rangle \otimes |0_\omega\rangle} \ldots C_{|1e2g;0_\omega\rangle \otimes |0_\omega\rangle} \ldots)$.

From this q-state there are no possibilities for the material system to emit one photon state. And chemistry can possibly henceforth proceed via quantum dissociation event. To see this let us construct two normalized linear superpositions $|\pm\rangle$:

$$|\pm\rangle = (0_{|1g2g\rangle \otimes |1_\omega\rangle} \ldots 0_{|1g2g;1_\omega\rangle \otimes |0_\omega\rangle} \ldots 1/\sqrt{2}\,_{|1g2e;0_\omega\rangle \otimes |0_\omega\rangle} \ldots \pm 1/\sqrt{2}\,_{|1e2g;0_\omega\rangle \otimes |0_\omega\rangle} \ldots) \rightarrow (15)$$

Coupling the states $|\pm\rangle$ one can get as a result either (16) or (17):

$$|+\rangle + |-\rangle \rightarrow (0_{|1g2g\rangle \otimes |1_\omega\rangle} \ldots 0_{|1g2g;1_\omega\rangle \otimes |0_\omega\rangle} \ldots 1_{|1g2e;0_\omega\rangle \otimes |0_\omega\rangle} \ldots 0_{|1e2g;0_\omega\rangle \otimes |0_\omega\rangle} \ldots) \quad (16)$$

$$|+\rangle - |-\rangle \rightarrow (0_{|1g2g\rangle \otimes |1_\omega\rangle} \ldots 0_{|1g2g;1_\omega\rangle \otimes |0_\omega\rangle} \ldots 0_{|1g2e;0_\omega\rangle \otimes |0_\omega\rangle} \ldots 1_{|1e2g;0_\omega\rangle \otimes |0_\omega\rangle} \ldots) \quad (17)$$

Thus, reading from labels either one could detect an exited state at one end and, fully independent of distance, there will be a ground state in the opposite direction. Take (17) and identify a detector to be label as D-X1; coincidence with the label amplitude $1_{|1e2g;0\omega>\otimes|0\omega>}$ permits transfer of a q-event to amplitude at base state $|1e2g;0\omega>\otimes|0\omega>$. As we construct theory and detector apparatus a conclusion follows: if at the antipode of D-X1 one sets up another detector it would not respond.

The conclusion may appear puzzling. For, in one way or another labels implicitly hide the R-parameter and since energies are R-independent the base state would share this character.

To actualize an R-dependence the bi-partite state should be made effective (semi-classic framework). Note that the interaction responsible for the recombination leading to (16) for example includes *simultaneous information on both ends*. The R-dependence will show up now if a q-event were to happen.

The q-state (17) *if detected* would also show separate base state response, namely, ground state at one end and excited states at the other that, by construction, are R-separated. If this is true, having detected excitation at one end the conclusion will be that the q-state shows a ground state system. *There would be no signal involved*, reflecting the nature of the entangled state. The transition between one- to bi-partite states results from q-interactions, and these ones do not fully belongs to the entangled state under scrutiny; it lies on a bridge between spaces. For large R, superluminal signals have no rational place; they are not even wrong. Entangled states once "activated" contain all necessary information.

**Dissociations competition**

Take a reactant system starting from ground state that may show two bond-dissociation possibilities measured by different dissociation limits: one at lowest energy than the other. The last one would show up first as product thence it may be followed by products from the higher energy one.

Now, take a 1-photon excitation at a chromophore partite with energy above the highest dissociation channel. The energy gaps are inverted with respect to this excited state level and one would expect to sense the products in inverse order.

For the photonic approach the response from the given materiality to external probes matters. The structural element enters as graph-label only. The spectroscopic idea replaces that of structure. In this sense, photonic and semi-classic frameworks results would complement each other as they target the same type of processes from opposite sides as it were. The procedure translated to laboratory information tells us something simple: if the excitation is detected at one end this mean that the q-state at the other end comes up as information and might be used with certainty. We do not need to carry on any further measure. There is no use for a *psychophysical* hypothesis (action), registering would occur in absence of an observer. In other words: there would be no classic decoherence; conclusion, materiality does not occupy energy label states.

**Information transfers**

In apprehending the meaning of wavefunctions, even from early stages, the information concept has occupied a central place. The initial representational content is not retained here. Response towards external probes connects to amplitudes that appear actually in control of the interaction. Amplitude variation of this kind can be activated either by a quantum event or by a q-interaction in Hilbert space (interaction operators).

Reading from labels is one basic aspect. It is there one can follow (construct) histories and permit telling stories. The state (6) concerns both q-fields, i.e. photon- and matter q-fields that varies at the same time; no free photon states are available once entangled and consequently there is no room for justifying independent semi-classic pictures.

Lab-processes related to changes involving many-partites states require entanglement first between partite states to come in (opening as it were) a responding q-space. Entangled states do not belong to partites states taken separately. These special states are added here to the base state vector, and, in so doing, the number of I-frames may go down by one unit. As a result, these entangled states play the fundamental role of linking spaces that otherwise are incommensurate.

This latter piece of information is central to apprehend processes seen from abstract space perspectives. *Spaces sustained by possibilities*. Physics without objects is the logical way to implement a q-theory for understanding chemical processes including photonics; one must become familiar with this new state of affairs before examining the type of description associated with chemical processes. D. Finkelstein noticed that q-systems ought to be seen as a plexus of q-processes (q-events for us) and not a plenum of q-objects. [22]

Yet, elementary materiality must be present. This materiality would be seen as information carriers to the extent they sustain extended q-states; at any rate, it is to be seen as *executive* presence. Electromagnetic energy actually is a carrier of energy quanta as well as quantized angular momenta. The classical representation with help from electric and magnetic fields is useful to construct coupling operators but badly miss q-events.

Along the line of information transfer, constructing bridges to semi-classic models would give supplementary dimension to modeling approaches. See for instance Berrada's et al. work on entanglement generation involving the model obtained from eq.(3) and a beam splitting device. [23]

Inclusion of SRT information, configuration space supported labels read: $|\mathbf{x},ict\rangle$ and $|\mathbf{p},iE/c\rangle$ with connecting function:[24]

$$\langle \mathbf{x},-ict \,|\, \mathbf{p},iE/c\rangle = (2\pi\hbar)^{-2} \exp(i(\mathbf{p}\cdot\mathbf{x} - Et)/\hbar) \qquad (18)$$

The arguments have in the realm of SRT kinematic meanings, while in abstract space they are employed as labels; c speed of light, E with dimension of energy.

Conservation laws e.g. energy, angular and linear momenta enter the photonic framework in a natural manner as soon as processes producing variation of the I-frames numbers. This advantage permits analyses of physical and chemical routes as illustrated above. To the extent biologic materials present similarities with photonics it is natural to use the present approach to think biologic processes in quantum physical terms as well. Yet they will never represent objects.

The photonic scheme gives an abstract view to chemical and physical processes. For biologically sustained processes it is well adapted too. Used to analyze semi-classic pictures a deeper understanding of natural processes would arise.

Opening access to spin triplet indicates a multi-photon mechanism regulated by conservation laws. Excitation wandering, for example, in any solar cell will perform well (including red shifts). The key is quantum coherence propagating via particular materiality. Coherence/decoherence would fix pathways via vertices able to prompt photon states emission. This sort of photon recycling would increase efficiency.[25]

**Discussion**

That a q-state emerges as image if *appropriately recorded* follows from the present photonic scheme. Tonomura's experiment for a double slit setup substantiate it;[16] for, initially collected q-events seem to stand for a random process, however, after gathering sufficient number of q-events, a supportive image develops in front of us corresponding more and more to what is named interference pattern. This is a sort of "impressionist" rendering of the q-state. However, even from the first "click" the q-state (amplitude set) is present as a possibility (covering all thinkable responses).

The above situation, if anything, shows a q-state is not an objet but it hangs somehow on sensitive surfaces revealing an image constructed out of q-events: q-energy and AM exchanges such is the reality of a q-state. There is no doubt that the materiality sustaining the q-state must arrive at the detecting surface, but *information* transported is richer than that a classical particle impact would convey.

Between a q-system and a q-detector there is "executive" energy *and* information exchange: q-events. These q-events actually suppress the standard view of decoherence. Now, one is in front of a physical process that mediates probing; not observers with their friends producing decoherence; this later is rather a non-computational foundation.

Activation events with one photon-state were explored using some simple examples.[26] Photon/matter states entanglement plays a key role hinging to q-events, a sort of mediating channel. Reversing direction the entangled state prompt for one photon event within a larger possibility bases. Lifetime would necessarily characterize the situation that has not yet included.

The opening of a spin-triplet state required an executive course starting from the nearest spin singlet excited electronic state. A base state affected by a zero-valued amplitude does not respond to an external probe. Use of the S-T energy gap injection as photon states with resonant frequency turns out to be a key to opening the channel towards setting a non-zero value amplitude at the triplet base state. A second resonant photon leads to a cascade accomplishing the triplet state activation, namely, non-zero amplitude. Measuring the process from the spin-singlet ground electronic state two units of angular momentum are necessary. The role of two S-T gap energy quanta legitimated angular momentum conservation.

Thus, injection of one energy quantum (photon) does not lead to a dynamical process. The role of low frequency radiation injected (shinned) onto the system will help support quantum dynamical processes. The important result is that no time evolution can be expected without presence of low-frequency EM energy quanta.

The possibility to produce multipartite states permits analyses of chemical processes. They can be seen related to amplitudes for entanglement related to forming/breaking sustainment, chemical bonds in chemist's language. This entanglement changes are to be taken as quantum processes not as mechanical ruptures/sewing of chemical bonds. Yet they can be related via q-events. For, q-events enter so that localization in laboratory space permits introducing laboratory space magnitudes. This is one of the advantages of the photonic scheme via I-frames. The q-events are effective energy and AM transfer processes linking the q-system to probing devices. [2,3,9,14,26]

Note, q-events and q-entanglements do not elicit computational algorithms in themselves. It is a matter of principle. On the other hand, registering becomes a resource employed to indicate relative distances and/or orientations. This is another feature of the present quantum approach.

A warning word is in place: a primal non-classical attribute of q-states unfortunately may acquire a pseudo-classical gloss if one imagines the superposition terms as a kind of classical interference of wave functions. Eradicate this type of imagination if a sensed perception of quantum physics is the goal.

The population idea fades away as the representational character of the theory is absent in the photonic scheme. Neither photons nor molecules are seen as objects. The energy exchanged between q-materials and photon fields corresponds to the quantity of energy assigned to the photon field. The reason explaining why we do refer to q-events is that no representational character is assigned to symbols. Q-event embodies a quantum of energy and angular momentum. The q-event can be spatially localized as well as timed. This type of process associated to changes in the number of I-frames thereby relates the event to particular real space localizations. The standard QM does not describe situations such as real energy/momentum exchange.

Taking Alain Aspect version of EPR (Einstein-Podolski-Rosen) experiment [15] use our language to examine this case. A one-partite system acts as source and "decompose" into a bi-partite state: each one with a "guiding" I-frame. This change in numbers of I-frames signals a q-event that in this case corresponds with a two photon-states emission. Conservation of linear momentum imposes photon states **k**-vectors to signal opposite directions. But the 1-partite states have internal quantum numbers e.g. eliciting polarization: i.e. an entangled state. Note that for laboratory cases the axis between **k**-vectors may be "tumbling" if the original I-frame or more properly that the photon states display higher space angular momentum.

Here we use our q-states (15), (16) and (17) to display quantum changes to get ready dissociation. The states |±> are entangled states, q-states; a q-event is model by either state (16) or (17). Because they are related to a one-component they will have

amplitudes affected by exp(i**kr**) and exp(-i**kr**). So that one can put detectors to sense excitation that now are identified either by amplitudes $1_{|1g2e;0_\omega> \otimes |0_\omega>}$ or $1_{|1e2g;0_\omega> \otimes |0_\omega>}$. But, reading from the labels one detect excitation say at a point while at the antipode probing there will be no excitation to detect. It is apparent that here there is no signal send. The quantum system has its own abstract space that has little to do with real space.

**A century later**. Almost a century has gone before we acknowledge that in quantum physics what is measured corresponds to quantum states sustained by material carriers and these latter express particle-states (not classical particles) via I-frames. There is no such a thing as wave-particle duality either;[15] this latter presents itself in our way of speaking only.[17] It is an ideology superposed to a banal mathematical analysis.

The I-frames opens also the opportunity to introduce the concept of NPP (nodal plane patterns). There are possibilities to take into account different types of symmetries without being stopped by the parametric dependent electronic wave functions. Also, prospects to construct bridges towards semi-classic schemes result therefrom.[4,27]

Naturally, it may happen that scattering and interference be present in classical wave phenomena as well; however, *discovering interference in a quantum setting does not logically imply a wave property in a classical sense*. Such conclusion would not be logically granted. And is quantum physically a non-sense as this latter actually is physics without objects.[9,14,18]

Both measured and measuring devices comprise quantum elements that might prompt for q-events. The quantum system can be subjected to q-interactions that do not imply detectable energy swap; this could be the case (at least in part) with a double-slit device or similar.

Thus, Quantum mechanics is not a representational scheme. It does not describe either particles or waves in a classical physics mode (not even analogically). This can be taken as the "message of the quantum": do not mix up the levels of analyses. In particular if no clear definitions of the concerned level are put forward. And this is rather usual in discussing "reality" in quantum physical settings. If one takes "reality" to be the "clicks" appearing in a particular experiment then, according to the present approach, they hide more than they uncover. In particular correlations induced by the wavefunction that the collected information would put in evidence once the image becomes apparent but, as possibilities, they are already there. So, all individual q-events (to recover our formulations) are not innately random. Yet, quantum physics says goodbye to reality is an unfortunate formulation. Photons are not "particles"; they are useful ideological construct helping in the communication business.

To speak of "Experiments with entangled pairs of particles" [28] already introduces some spurious elements. For, what one would be planning with these semi-classic experiments is to probe quantum states sustained by an elementary materiality (named particles here). The q-event is then transformed into particle localization without even noticing that what are being probed concern q-states and not particle trajectories.

The crux of Zeilinger and coworkers study is the analysis of the semi-classic scheme used to scrutinize data *the results lend strong support to the view that any future* *extension of* (semi-classic) *quantum theory that is in agreement with experiments must abandon certain features of realistic descriptions*. From the present viewpoint this conclusion seems most useful.

A clear distinction of abstract and laboratory levels must be established always; the semi-classic quantum mechanical mode should not be mixed up with both abstract and standard Quantum Mechanics. Initially, immersed in a still dominating classical worldview quantum physics symbols for particle positions were mapped on to linear self-adjoint operators while keeping classical overtones so that taking a "limit" h$\rightarrow$0 one could recover classical mechanics counter parts. But such a limit is illusory for Planck constant sanctions a fundamental different form of interactions: "quantumness" in the exchange of energy and information between laboratory q-systems, classical trajectories are meaningless.

Inclusion of radiation q-field opens possibilities to apprehend the idea of sustainment. The role of vacuum radiation base states keep information on histories reflected by elementary materiality via coherent q-states. Vacuum role emerges once q-radiation is "shinned" onto the system. Entanglement processes then open the partite contributors to q-interaction and q-events.

**Acknowledgement**

The author is indebted to Prof. E. Ludeña for cogent and enlightening discussions.